\documentstyle[12pt]{article}
\newtheorem{deff}{Definition}[section]

\def\la{\leftarrow}
\def\ra{\rightarrow}

\def\bd{\noindent\bf}
\def\sbd{\vspace{8pt}\noindent\bf}

\newbox\tempa
\newbox\tempb
\newdimen\tempc
\def\mud#1{\hfil $\displaystyle{\mathstrut #1}$\hfil}
\def\rig#1{\hfil $\displaystyle{#1}$}
\def\irulehelp#1#2#3{\setbox\tempa=\hbox{$\displaystyle{\mathstrut #2}$}%
		        \setbox\tempb=\vbox{\halign{##\cr
	\mud{#1}\cr
	\noalign{\vskip\the\lineskip}%
	\noalign{\hrule height 0pt}%
	\rig{\vbox to 0pt{\vss\hbox to 0pt{${\; #3}$\hss}\vss}}\cr
	\noalign{\hrule}%
	\noalign{\vskip\the\lineskip}%
	\mud{\copy\tempa}\cr}}%
		      \tempc=\wd\tempb
		      \advance\tempc by \wd\tempa
		      \divide\tempc by 2 }
\def\irule#1#2#3{{\irulehelp{#1}{#2}{#3}%
		     \hbox to \wd\tempa{\hss \box\tempb \hss}}}

\begin{document}
\begin{center}
{\Large{\bf The Undecidability of Pattern Matching in Calculi where 
Primitive Recursive Functions are Representable}}\\[20pt]
{\bf Gilles Dowek}\\[10pt]
{\bf INRIA}\def\thefootnote{\fnsymbol{footnote}}\footnote[1]{B.P. 105, 78153
Le Chesnay CEDEX, France. dowek@margaux.inria.fr}\footnote[2]{This research
was partly supported by ESPRIT Basic Research Action ``Logical Frameworks''.}
\def\thefootnote{\arabic{footnote}}
\\[15pt]
\end{center}

We prove that the pattern matching problem is undecidable in 
polymorphic $\lambda$-calculi (as Girard's system $F$ \cite{Girard} 
\cite{Giraflor}) and calculi supporting inductive types 
(as G\"odel's system $T$ \cite{Godel} \cite{Giraflor})
by reducing Hilbert's tenth problem to it.
More generally pattern matching is undecidable in all the calculi in which 
primitive recursive functions can be fairly represented in a precised sense.

\section*{Introduction}

The {\it higher order matching} problem in a typed $\lambda$-calculus is the
problem of determining whether a term is an instance of another i.e. to 
solve the equation $a = b$ where $a$ and $b$ are terms and $b$ is ground. The 
decidability of pattern matching in simply typed $\lambda$-calculus is still 
an open problem.

Extensions of simply $\lambda$-calculus are obtained by adding 
dependent types, polymorphism, type constructors and inductive types.
In \cite{cras} we have proved that pattern matching is undecidable in 
$\lambda$-calculi with dependent types or type constructors. We prove in this
note that pattern matching is also undecidable in polymorphic 
$\lambda$-calculi (as Girard's System $F$ \cite{Girard} \cite{Giraflor}) and 
in $\lambda$-calculi supporting inductive types (as G\"{o}del's System $T$
\cite{Godel} \cite{Giraflor}).
More generaly a $\lambda$-calculus
cannot at the same time be sufficiently expressive to represent primitive 
recursive functions and let pattern matching be decidable.

\section{Girard's System $F$}

We use the definition of system $F$ and the notations of \cite{Barendregt} 
except that we write $Prop$ instead of $*$ and $t[x \la t']$ for the term 
obtained by substituting the term $t'$ for the variable $x$ in the term $t$. 

{\sbd Definition:} Syntax

$$T~::=~Prop~|~x~|~(T~T)~|~\lambda x:T.T~|~\Pi x:T.T$$

The notation $T \ra T'$ is an abbreviation for $\Pi x:T.T'$ when $x$ has no
occurrence in $T'$.

{\sbd Definition:} Context

A {\it context} is a list of pairs $<x,T>$ (written $x:T$) where $x$ is a 
variable and $T$ a term. 

{\sbd Definition:} Typing Rules

We define inductively two judgements: {\it $\Gamma$ is well-formed} and 
{\it $t$ has type $T$ in $\Gamma$} ($\Gamma \vdash t:T$) where $\Gamma$ is
a context and $t$ and $T$ are terms.

$$\irule{} 
           {[~]~\mbox{well-formed}}
           {}$$

$$\irule{\Gamma \vdash T:Prop} 
           {\Gamma[x:T]~\mbox{well-formed}}
           {}$$

$$\irule{\Gamma~\mbox{well-formed}} 
           {\Gamma[x:Prop]~\mbox{well-formed}}
           {}$$

$$\irule{\Gamma~\mbox{well-formed}~~x:T \in \Gamma} 
           {\Gamma \vdash x:T}
           {}$$

$$\irule{\Gamma \vdash T:Prop~~\Gamma[x:T] \vdash T':Prop}
           {\Gamma \vdash \Pi x:T.T':Prop}
           {}$$

$$\irule{\Gamma[x:Prop] \vdash T:Prop}
           {\Gamma \vdash \Pi x:Prop.T:Prop}
           {}$$

$$\irule{\Gamma \vdash T:Prop~~\Gamma [x:T] \vdash T':Prop~~
\Gamma[x:T] \vdash t:T'} 
           {\Gamma \vdash \lambda x:T.t:\Pi x:T.T'}
           {}$$

$$\irule{\Gamma [x:Prop] \vdash T:Prop~~\Gamma[x:Prop] \vdash t:T} 
           {\Gamma \vdash \lambda x:Prop.t:\Pi x:Prop.T}
           {}$$

$$\irule{\Gamma \vdash t:(x:T)T'~~\Gamma \vdash t':T}
           {\Gamma \vdash (t~t'):T'[x \la t']}
           {}$$

{\sbd Definition:} $\beta$-reduction and $\beta$-equivalence

The $\beta$-reduction (in one step) ($\rhd$) is the smallest 
relation compatible with term structure that verifies:
$$(\lambda x:T.t~u) \rhd t[x \la u]$$

The $\beta$-reduction relation ($\rhd^{*}$) is the reflexive-transitive
closure of the relation $\rhd$ and the $\beta$-equivalence ($\equiv$) 
is the reflexive-symmetric-transitive closure of the relation $\rhd$.

{\sbd Definition:} Normal Term

A term $t$ is said to be {\it normal} if there exists no term $u$ such 
that $t \rhd u$.

{\sbd Remark:} The proof given in this note also works if we consider 
$\eta$-reduction too. 

{\sbd Theorem:}
The reduction on well-typed terms is strongly normalizable and confluent, i.e.
all the reduction sequences issued from a well-typed term $t$ are finite and 
if $u$ and $u'$ are normal terms such that $t \rhd^{*} u$ and
$t \rhd^{*} u'$ then $u = u'$.

{\bd Proof:} See \cite{Girard} \cite{Giraflor} for the $\beta$-reduction
and \cite{Gallier} \cite{Geuvers} \cite{Salvesen} for the generalization to 
$\beta \eta$-reduction.

{\sbd Proposition:}
Let $t$ be a normal well-typed term, $t$ is either an abstraction, a product 
or an atomic term i.e. a term of the form $(w~c_{1}~...~c_{p})$ where $w$ is 
a variable or a sort.

{\bd Proof:} If the term $t$ is neither an abstraction nor a product then it 
can be written in a unique way $t = (w~c_{1}~...~c_{p})$ where $w$ is not an 
application. 
The term $w$ is not a product (if $p \neq 0$ because a product is of type 
$s$ for some sort $s$ and therefore cannot be applied and if $p = 0$ because
$t$ is not a product).
It is not an abstraction (if $p \neq 0$ because $t$ is in normal form
and if $p = 0$ because $u$ is not an abstraction). It 
is therefore a variable or a sort.

{\sbd Definition:}
We let $Nat = \Pi P:Prop. P \ra (P \ra P) \ra P$
and for every natural number $n$, $\overline{n}$ be the Church natural
representing $n$:
$$\overline{n} = 
\lambda P:Prop. \lambda x:P. \lambda f:P \ra P.(f~...~(f~x)~...~)
~\mbox{($n$ times)}$$

{\sbd Proposition:}
For every primitive recursive function $f$ of arity $n$,
there exists in system $F$ a term $t$ of type $Nat \ra ... \ra Nat \ra Nat$ 
such that if $a_{1}, ..., a_{n}$ are natural numbers, 
then:
$$(t~\overline{a_{1}}~...~\overline{a_{n}}) = \overline{(f~a_{1}~...~a_{n})}$$
Moreover the term $t$ can be effectively constructed from the definition of 
$f$. The term $t$ is said to {\it represents} the function $f$.

{\bd Proof:} See \cite{Girard} \cite{Giraflor}.

\section{The Undecidability of Primitive Recursive Equations}

Let us recall some well-known facts about primitive recursive functions.

{\sbd Proposition:}
The following functions are primitive recursive:

$\bullet$ addition and multiplication,

$\bullet$
the function $Equal$ such that $(Equal~x~y) = 0$ if $x = y$ and 
$(Equal~x~y) = 1$ otherwise,

$\bullet$ the function $\alpha$ such that $(\alpha~x~n)$ is the exponent of the
$n^{th}$ prime number in the prime decomposition of $x$.

{\sbd Proposition:}
For every finite sequence of natural numbers $a_{1}, ..., a_{n}$, 
there exists an natural number $x$ such that for every $i$, $1 \leq i \leq n$,
$a_{i} = (\alpha~x~i)$. 

{\bd Proof:} We take $x = \Pi_{i=1}^{n} p_{n}^{a_{n}}$ where $p_{n}$ is the
$n^{th}$ prime number.

{\sbd Proposition:} 
There is no effective method that decides if, given the definition of a 
primitive recursive function $f$, the equation $(f~x_{1}~...~x_{n}) = 0$ has a
solution.

{\bd Proof:} 
We reduce Hilbert's tenth problem \cite{Davis} to this one.
Let $(P~x_{1}~...~x_{n})$ and $(Q~x_{1}~...~x_{n})$ two polynomials.
Let us define the function $f$ as:
$$(f~x_{1}~...~x_{n}) = (Equal~(P~x_{1}~...~x_{n})~(Q~x_{1}~...~x_{n}))$$
The equation
$(f~x_{1}~...~x_{n}) = 0$ has a solution if and only if
$(P~x_{1}~...~x_{n}) = (Q~x_{1}~...~x_{n})$ also has one.

{\sbd Remark:}
In the previous proposition we can restrict ourselves to equations with only
one variable by taking:
$$(f~x) = (Equal~(P~(\alpha~x~1)~...~(\alpha~x~n))
~(Q~(\alpha~x~1)~...~(\alpha~x~n)))$$

\section{The Undecidability of Pattern Matching in Girard's system $F$}

{\sbd Definition:}
A {\it matching problem on one natural variable} is a pair of terms $<a,b>$ 
such that 
$a$ is well-typed in the context $[x:Nat]$ and $b$ is well-typed in the
empty context. A solution of such a problem is a pair $<\gamma,u>$ such that
$\gamma$ is a well-formed context and $u$ a term well-typed of type $Nat$ in 
the context $\gamma$ such that $a[x \la u]$ and $b$ have the same
normal form (these two terms are well-typed in the context $\gamma$).

{\sbd Remark:} Although $a$ may have only $x$ as free variable and $b$ does not
have any, there is no restriction on the free variables of the term 
$u$ since $\gamma$ is an arbitrary well-formed context.

{\sbd Proposition:}
Let $\Gamma$ be a context and $t$ a normal term well-typed in $\Gamma$ of type
$Nat$ such that the normal form of $(t~Nat~\overline{0}~\lambda y:Nat.y)$ is 
$\overline{0}$ then the term $t$ is a Church natural.

{\bd Proof:} 
Let us consider the context $\Gamma' = [P:Prop; x:P; f:P \ra P]$. Let the term
$u$ be the normal form of $(t~P~x~f)$. The term $u$ has type $P$ in $\Gamma'$.
We have:
$$(t~Nat~\overline{0}~\lambda y:Nat.y) \equiv \overline{0}$$
so:
$$u[P \la Nat, x \la \overline{0}, f \la \lambda y:Nat.y] \equiv \overline{0}$$
We prove by induction over the structure of $u$ that every normal term $u$ of 
type $P$ in the context $\Gamma'$ such that the normal form of 
$u[P \la Nat, x \la \overline{0}, f \la \lambda y:Nat.y]$ is $\overline{0}$
has the form $u = (f~...~(f~x)~...~)$.

The term $u$ has type $P$ so it is neither an abstraction nor a product.
It is thus an atomic term $(w~c_{1}~...~c_{p})$. 
If $w$ is different from $P$, $f$ and $x$ then the normal form of the term
$u[P \la Nat, x \la \overline{0}, f \la \lambda y:Nat.y]$ is also atomic with 
head $w$ and thus is different from $\overline{0}$.
So the variable $w$ is among $P$, $f$ and $x$. It is not the variable $P$
because we would have $p = 0$ and the normal form of 
$u[P \la Nat, x \la \overline{0}, f \la \lambda y:Nat.y]$ would be the term
$Nat$ which is not $\overline{0}$, so it is either $x$ or $f$.

If $w = x$ then $p = 0$ so $u = x$ has the required form.
If $w = f$ then $p = 1$, $u= (f~u')$.
The term $u[P \la Nat, x \la \overline{0}, f \la \lambda y:Nat.y]$ reduces to
$u'[P \la Nat, x \la \overline{0}, f \la \lambda y:Nat.y]$, so the normal form 
of this term is $\overline{0}$.
Thus, by induction hypothesis, we have $u' = (f~...~(f~x)~...~)$ 
and $u = (f~(f~...~(f~x)~...~))$ has the required form.

At last since the normal form of the term $(t~P~x~f)$ is $(f~...~(f~x)~...~)$
and this term has not the form $(v~f)$ with $v$ normal we have
$t = \lambda P:Prop. \lambda x:P.\lambda f:P \ra P.(f~...~(f~x)~...~)$.

{\sbd Theorem:} There is no effective method that decides if a matching problem
on one natural variable in system $F$ has a solution.

{\bd Proof:} Let $f$ be an unary primitive recursive $f$, we build a 
matching problem on one natural variable $<a,b>$ that has a solution if and 
only if $f$ takes the value $0$.
Let $t$ be a term representing the function $f$ and $Pair$ be the term:
$$Pair = \lambda x:Nat. \lambda y:Nat. \lambda g:Nat \ra Nat \ra Nat.(g~x~y)$$
Let:
$$a = (Pair~(x~Nat~\overline{0}~\lambda y:Nat.y)~(t~x))$$
$$b = (Pair~\overline{0}~\overline{0})$$
Let $n$ be a natural number such that $(f~n) = 0$, 
the pair $<[~],\overline{n}>$ is a solution of the matching problem $<a,b>$.
Conversely, let $<\gamma,u>$ be a solution of the matching problem $<a,b>$, 
the normal form of the term $(u~Nat~\overline{0}~\lambda y:Nat.y)$ is 
$\overline{0}$ and the normal form of $(t~u)$ is $\overline{0}$.
Thus the normal form of $u$ is a Church natural $\overline{n}$ and $(f~n) = 0$.

{\sbd Remark:} In \cite{matching} we have developed a more general notion
of matching problem ans made a distinction between {\it universal variables}
that cannot be instanciated by a substitution and {\it existential variables}
that can be instanciated by a substitution. 
We have also defined a notion of {\it order} of a type $T$ in a context 
$\Gamma$:
\begin{itemize}
\item if $T$ is atomic, $T = (w~c_{1}~...~c_{n})$ then if $w$ is an universal 
variable then $o(T) = 1$, if $w$ is an existential variable then 
$o(T) = \infty$ and if $w$ is a sort then $o(T) = 2$,
\item if $T = \Pi y:U.V$ then $o(T) = max \{1 + u, v\}$
where $u$ is the order of $U$ in $\Gamma$ and $v$ is the order of $V$ in 
$\Gamma [y:U]$ letting $y$ be an existential variable
(with the usual conventions $n + \infty = \infty$ and 
$max \{n,\infty\} = \infty$).
\end{itemize}

We have proved in \cite{matching} that second order matching was decidable
in all the systems of the cube of type systems \cite{Barendregt} including 
system $F$. 
Since the order of $Nat$ is infinite, the problems considered in this note are
of infinite order. So the problem of decidability of pattern matching in system
$F$ with only finite order variables is left open. 
Since restricting the order of variables to finite order prohibits the use of
polymorphism, this problem seems related to the problem of pattern matching in
simply typed $\lambda$-calculus.

\section{The Undecidability of Pattern Matching in G\"{o}del's System $T$}

{\sbd Definition:}
G\"{o}del's System $T$ \cite{Godel} \cite{Giraflor} is an extension of simply
typed $\lambda$-calculus in which:
\begin{itemize}
\item{there are a primitive type $Nat$ and primitive symbols 
$O:Nat$ and $S:Nat \ra Nat$,}
\item{for each type $T$, there is a primitive symbol $R_{T}$ 
(called the recursor of type $T$) of type 
$T \ra (Nat \ra T \ra T) \ra Nat \ra T$,}
\item{reduction is extended by the rules:
$$(R_{T}~a~b~O) \rhd a$$
$$(R_{T}~a~b~(S~x)) \rhd (b~x~(R_{T}~a~b~x))$$}
\end{itemize}

{\sbd Theorem:}
The reduction on well-typed terms is strongly normalizable and confluent.

{\bd Proof:} See \cite{Godel} \cite{Giraflor}.

{\sbd Remark:} Usually $\eta$-reduction is not considered in system $T$.
The proof given here also works if we consider $\eta$-reduction too, 
provided that the reduction relation is strongly normalizable and confluent.

{\sbd Proposition:}
Let $\overline{n}$ be the term $(S~...~(S~O)~...~)$ ($n$ times).
For every primitive recursive function $f$ of arity $n$,
there exists in system $T$ a term $t$ of type $Nat \ra ... \ra Nat \ra Nat$ 
representing the function $f$, moreover the term $t$ can be effectively 
constructed from the definition of $f$.

{\bd Proof:} See \cite{Godel} \cite{Giraflor}.

{\sbd Proposition:}
In G\"{o}del's system $T$, let $t$ be a normal
term of type $Nat$ such that the normal form 
of $(R_{Nat}~O~\lambda y:Nat. \lambda z:Nat.z~t)$ is $O$ then $t$ has the form 
$(S~...~(S~O)~...~)$.

{\bd Proof:} By induction over the structure of $t$. 
The term $t$ has type $Nat$ so it not an abstraction, since it is normal it is
an atomic term $(w~c_{1}~...~c_{p})$.
If $w$ is different from $O$ and $S$ then the term 
$(R_{Nat}~O~\lambda y:Nat. \lambda z:Nat.z~t)$ is normal and is different from 
$O$. So the variable $w$ is either $O$ or $S$. 

If $w = O$ then $p = 0$ so $t = O$ has the required
form. If $w = S$ then $p = 1$, $t = (S~t')$.
The term $(R_{Nat}~O~\lambda y:Nat. \lambda z:Nat.z~t)$ reduces to
$(R_{Nat}~O~\lambda y:Nat. \lambda z:Nat.z~t')$, so the normal form 
of this term is $O$. Thus, by induction hypothesis, 
$t' = (S~...~(S~O)~...~)$ and $t = (S~(S~...~(S~O)~...~))$ has the required 
form.

{\sbd Theorem:} There is no effective method that decides if a matching problem
on one natural variable in system $T$ has a solution.

{\bd Proof:} The proof is the same as the one for system $F$,
except that we replace the term 
$(x~Nat~\overline{0}~\lambda y:Nat.y)$ by the term
$(R_{Nat}~O~\lambda y:Nat. \lambda z:Nat.z~x)$.

{\sbd Remark:}
In system $T$ the type $Nat$ is primitive, so even first-order pattern 
matching is undecidable.

\section*{Conclusion}

In this note we have proved the undecidability of pattern matching in
system $F$ and system $T$. The proofs given here generalize to all the 
polymorphic systems of the cube of typed $\lambda$-calculi \cite{Barendregt},
to all the systems of this cube extended by inductive types 
\cite{CoqPau} and to Martin-L\"{o}f's Type Theory \cite{MartinLof}.
More generally if we say that primitive recursive functions can be 
{\it fairly} represented in a typed $\lambda$-calculus 
when these functions can be represented and there exists a term $t$ of type 
$Nat \ra Nat$ such that if $u$ is a term of type $Nat$ then the term
$(t~u)$ reduces to $\overline{0}$ if and only if $u$ 
represents an integer, then pattern matching is undecidable in all the systems
in which primitive recursive functions can be fairly represented.

In \cite{cras} we have proved the undecidability of pattern 
matching in calculi with dependent types and type constructors. Pattern
matching is therefore undecidable in seven calculi of the cube of typed 
$\lambda$-calculi \cite{Barendregt}. The problem of the decidability of 
pattern matching in simply typed $\lambda$-calculus is left open. This
problem is conjectured decidable in \cite{Huet76}.

\section*{Acknowledgments}

The author would like to thank Amy Felty, Serge Grigorieff, G\'erard Huet and
Christine Paulin for their help in the preparation of this note.

\end{document}